\documentclass[letterpaper]{mn2e}

\usepackage{epsfig,psfig}

\newcommand{\mnras}{MNRAS}
\newcommand{\prd}{Phys. Rev. D}
\newcommand{\aap}{AAP}
\newcommand{\apj}{ApJ}
\newcommand{\be}{\begin{equation}}
\newcommand{\ee}{\end{equation}}
\newcommand{\ba}{\begin{eqnarray}}
\newcommand{\ea}{\end{eqnarray}}

\newcommand{\nn}{\nonumber \\}
\newcommand{\x}{\mbox{\boldmath $x$}}

\newcommand{\s}{\mbox{\boldmath $s$}}

\newcommand{\C}{\mbox{\boldmath $C$}}
\newcommand{\Pp}{\mbox{\boldmath $P$}}
\newcommand{\Ct}{\mbox{\boldmath ${\tilde C}$}}

\newcommand{\D}{\mbox{\boldmath $D$}}

\newcommand{\mub}{\mbox{\boldmath $\mu$}}
\newcommand{\Pb}{\mbox{\boldmath $P$}}

\newcommand{\kk}{\mbox{\boldmath $k$}}

\newcommand{\thetab}{\mbox{\boldmath $\theta$}}

\newcommand{\n}{\mbox{\boldmath $n$}}

\newcommand{\rgl}{\rangle}
\newcommand{\lgl}{\langle}
\newcommand{\de}{\partial}

\newcommand{\Tr}{{\rm Tr}\,}
\newcommand{\half}{\frac{1}{2}}
\newcommand{\psib}{\mbox{\boldmath $\psi$}}

\newcommand{\calL}{\mbox{${\cal L}$}}

\def\gs{\mathrel{\raise1.16pt\hbox{$>$}\kern-7.0pt %
\lower3.06pt\hbox{{$\scriptstyle \sim$}}}}         %
\def\ls{\mathrel{\raise1.16pt\hbox{$<$}\kern-7.0pt %
\lower3.06pt\hbox{{$\scriptstyle \sim$}}}}         %

\title[Path Integral Marginalization]{
Path Integral Marginalization for Cosmology: \\
Scale Dependent Galaxy Bias \& Intrinsic Alignments}

\author[T. D. Kitching \& A. N. Taylor]
       {T. D. Kitching\thanks{tdk@roe.ac.uk}
\& A. N. Taylor\thanks{ant@roe.ac.uk}\\
Scottish Universities Physics Alliance (SUPA), Institute for
Astronomy, School of Physics and Astronomy,\\ University of
Edinburgh, Royal Observatory, Blackford Hill, Edinburgh, EH9 3HJ,
U.K.}

\date{}

\pagerange{\pageref{firstpage}--\pageref{lastpage}}

\pubyear{2010}

\begin{document}

\maketitle

\label{firstpage}

\begin{abstract}
We present a path-integral likelihood formalism that extends
parameterized likelihood analyses to include continuous
functions.
The method finds the maximum likelihood point in function-space, and
marginalizes over all possible functions, under the
assumption of a Gaussian-distributed function-space.
We apply our method to the problem of removing unknown systematic
functions in two topical problems for dark energy research :
scale-dependent galaxy bias in redshift surveys; 
and galaxy intrinsic alignments in cosmic shear surveys. 
We find that scale-dependent galaxy bias
will degrade information on cosmological parameters unless the
fractional variance in the bias function is known to $10\%$. 
Measuring and removing intrinsic alignments from cosmic
shear surveys with a flat-prior can reduce the dark energy
Figure-of-Merit by $20\%$, however provided that the
scale and redshift-dependence is known to better than $10\%$ with
a Gaussian-prior, the dark energy Figure-of-Merit can be 
enhanced by a factor of two with no extra assumptions.
\end{abstract}


\section{Introduction}

The Standard Cosmological Model is a phenomenological model
containing two unknown components, dark energy and dark matter,
and two untested assumptions, general relativity on cosmic scales
and cosmological inflation. To understand this model at a deeper
level, by testing the evolution of dark energy, the nature of the
dark matter, signatures of inflation and possible deviations from
Einstein gravity, we need highly sensitive cosmological probes
that must constrain an expanded set of cosmological parameters to
(sub-)percent accuracy. In such a scenario the correct assessment
of systematic effects is of critical importance. 

To account for systematic effects it is common to use a parameterized 
description of the effect where each nuisance parameter is calibrated 
to some accuracy and removed. However, there are many cases where characterization of the
systematic effect by a parameterized function is not
well-justified, and in some cases the systematic is
effectively an unknown function. The assumption of a given
functional form can lead to a bias in measured cosmological
parameters, or at least an underestimate of the contribution to
errors due to uncertainty in the systematic function.

In Taylor \&
Kitching (2010) we described a new method which allowed the
analysis of large-dimensional parameter-spaces composed of
cosmological parameters and nuisance parameters describing
residual systematics, in both calibrated (Gaussian prior) and
self-calibrating (flat-prior) regimes.
Here we present a new method that can account for systematics in a
model-free way such that the impact of every possible function on
the cosmological signal is accounted for. To do this we introduce
a path-integral marginalization approach in which we maximize the
likelihood and  marginalize over a space of functions rather than
parameters. We construct a likelihood functional in parameter
space that accounts for the impact of all known systematic
functions. The function-space may be either weighted with a
flat prior about a fiducial function, where the data itself is
used to measure and remove the systematic effect, or if there is
external data weighted with a Gaussian prior, with a variance that
is zero when the systematic function is known.

We derive an expression for the Fisher matrix that accounts for
functional marginalization, and apply this method to two test
cases in cosmology in which the cosmological signal is
contaminated by a systematic effect with a poorly understood
functional behaviour. The first is scale-dependent galaxy bias
(e.g.,  Peacock \& Smith, 2000; Conway et al., 2005; Hamann et
al., 2008; Cresswell \& Percival, 2009) in redshift surveys, an
unknown function that relates the underlying dark matter distribution
to the galaxy distribution. 
The second is cosmic
shear intrinsic alignments (e.g. Brown et al., 2002 ; Crittenden
et al , 2001, Catelan et al,, 2001; Heymans et al., 2004; Hirata
\& Seljak, 2004; Kitching et al., 2007; Bridle \& King, 2007;
Joachimi \& Bridle, 2009; Joachimi \& Schneider, 2008, 2009;
Mandelbaum et al., 2009; Schneider \& Bridle, 2009; Kirk et al.,
2010) in weak lensing surveys, a contaminant with a complex 
astrophysical function that relates the local density to the
ellipticity of background galaxy images. In both cases we show how 
cosmological parameter
constraints are expected to degrade when we marginalize over all
systematic functions.

This work is an extension of the methods of  Taylor \& Kitching
(2010) in which a formalism for analytic marginalization was
developed for a finite number of parameters, and is related to the
Functional Form-Filling method of Kitching et al. (2009) in which
a numerical approach to assessing the effects of unknown functions
was presented.

This paper is organized as follows, in Section \ref{Path Integral
  Marginalization} we introduce path integral marginalization, in
Sections \ref{Scale-dependent, stochastic galaxy bias} and
\ref{Cosmic Shear Intrinsic Alignments} we apply the method to galaxy bias and
cosmic shear intrinsic alignments respectively. In Section
\ref{Conclusions} we present conclusions. Throughout we have put
technical detail in a series of Appendices.

\section{Path-Integral Marginalization}
\label{Path Integral Marginalization}

We consider an arbitrary likelihood functional, $L[\thetab,
\psib(\x|\thetab)]$, which is a function of a finite number of
cosmological parameters of interest, $\thetab$, and a finite
number of continuous nuisance functions, $\psi_\alpha(x|\thetab)$,
labelled by Greek subscripts. The nuisance functions could depend on
position $x=\x$, or wavevector $x=\kk$ if we are working in Fourier space.
The nuisance functions may themselves also depend on the cosmological
parameters. For convenience we shall work with the log-likelihood
functional;
\be
\calL [\thetab, \psib(x|\thetab)]
= - 2 \ln L[\thetab, \psib(x|\thetab)].
\ee
Expanding the log-likelihood functional to second order around an
arbitrary nuisance function we find
 \ba
       \label{funcexpansion}
    \calL &=&  \calL_0 +
    \int \! dx' \, \delta \psi_\alpha(x')
     \frac{\delta \calL_0}{\delta \psi_\alpha(x')} \\
        & + & \!\!\!\!
    \half \int \! dx' dx'' \, \delta \psi_\alpha(x')
     \left[\frac{\delta^2 \calL_0}{
     \delta \psi_\alpha(x')\delta \psi_\beta(x'')}\right]
      \delta \psi_\beta(x''). \nonumber
 \ea
This can be minimized in function-space with the solution
 \be
    \delta \psi_\alpha(x|\thetab)= - \int \! dx' \,
    \frac{\delta \calL_0}{\delta \psi_\beta(x')}
    \left[\frac{\delta^2 \calL_0}{
     \delta \psi_\alpha(x)\delta \psi_\beta(x')}\right]^{-1},
     \label{maxlikesol}
 \ee
where $\delta \psi_\alpha(x|\thetab)$ is the displacement in
function-space from the fiducial function and the
maximum-likelihood solution, and we have assumed the inverse of
the second functional derivative of the log-likelihood exists.

\subsection{Flat Priors}

Marginalizing over all paths for the functions, assuming a
flat-prior, with boundaries sufficiently far away they do not
contribute to the integral, we find
 \ba
 \label{marglike}
    \calL \!\!\!\!&=& \!\!\!\! \calL_0
    + \int dx' dx'' \Tr \ln\left[\frac{\delta^2 \calL_0}{
     \delta \psi_\alpha(x')\delta \psi_\beta(x'')}\right] \\
     \!\!\!\! & - & \!\!\!\!
  \half \int dx' dx''  \frac{\delta \calL_0}{\delta \psi_\alpha(x')}
    \!
     \left[\frac{\delta^2 \calL_0}{
     \delta \psi_\alpha(x')\delta \psi_\beta(x'')}\right]^{-1} \!
      \!\!\frac{\delta \calL_0}{\delta \psi_\beta(x'')},\nonumber   
 \ea
where we have omitted an unimportant constant.

The first and third terms in this expression represent the maximum
log-likelihood value, which can be seen by substituting the
solution  for the nuisance function $\psib(x|\thetab)$ given by
equation (\ref{maxlikesol}) into equation (\ref{funcexpansion}).
As we always find the maximum-likelihood solution for the nuisance
functions, the marginalized likelihood is independent of the
fiducial nuisance functions, $\psib(x|\thetab)$ in the Gaussian
approximation. The second term in equation (\ref{marglike}) is the
width of the likelihood function.

In Taylor \& Kitching (2010) we showed that in the case of
marginalization over discrete nuisance parameters the Fisher
information is preserved about the cosmological parameters of
interest. This result holds in the case of continuous functions,
$\psib(x|\thetab)$.

\subsection{Gaussian Priors}

If the nuisance functions are constrained by another, independent
experiment we can include this constraint as a Gaussian prior on
 $\psib(x|\thetab)$, with covariance matrix
 \be
        C_{\alpha\beta}(x,x') =
        \lgl \delta \psi_\alpha(x|\thetab) \,\delta \psi_\beta
        (x'|\thetab)\rgl.
 \ee
Including this constraint in the marginalization path integral we
find
 \ba
\label{lll}
    \calL \!\!\!\!&=& \!\!\!\! \calL_0 -
  \half \int dx' dx''  \frac{\delta \calL_0}{\delta \psi_\alpha(x')}
     M_{\alpha \beta}^{-1}(x',x'') \!
      \frac{\delta \calL_0}{\delta \psi_\beta(x'')} \nn
      & &
      + \int dx' dx'' \Tr \ln \half
      C_{\alpha \delta}(x',x'') M_{\delta \beta}(x',x''),
     \label{marglikeconstrained}
 \ea
where
 \be
    M_{\alpha \beta}(x',x'')= \frac{\delta^2 \calL_0}{
     \delta \psi_\alpha(x')\delta \psi_\beta(x'')} +
     2 C_{\alpha\beta}^{-1}(x',x'')
 \ee
and we assume the matrix $M_{\alpha \beta}$ is invertible. The
flat-prior, which we have assumed is non-zero over a finite but
sufficiently large region, is not a limiting case of the
Gaussian-prior. When the covariance matrix of the Gaussian-prior
goes to infinity, the prior distribution becomes arbitrary close
to zero everywhere, and the infinite wings of the prior 
suppress information.

\subsection{Gaussian Likelihoods with Systematics in the Mean}

A common distribution for cosmological datasets is that of a
multivariate Gaussian where, as discussed in Taylor \& Kitching
(2010), the effects of cosmological parameters and nuisance
parameters can appear in both the mean and/or the covariance. The
likelihood Gaussian function is given by
 \be
    \calL_0 = \Delta \D \C^{-1} \Delta \D^t + \Tr \ln \C,
 \ee
where $\Delta \D = \D - \mub$ is the variation of the data around
its mean value, $\mub = \mub[\thetab, \psib(\x|\thetab)]$, and
$\C[\theta,\psib(x|\thetab)] = \lgl \Delta \D \Delta \D^\dag\rgl$
is the data covariance matrix.

For simplicity we shall assume the nuisance functions and
cosmological parameters are only in the mean term, $\mub$, however 
as shown in Taylor \& Kitching (2010) results can be generalized for
nuisance functions in the covariance matrix. We assume the
curvature of the likelihood in function-space can be approximated
by its ensemble average, so that 
 \ba
 \label{llp}
    F_{\alpha\beta}(x,x') &=&
    \half \left\lgl \frac{\delta^2 \calL_0(x,x')}{
    \delta \psi_\delta(x)\delta \psi_\beta(x')}
    \right\rgl \nn
    &=&
    \Tr \,
    \C^{-1} \frac{\delta \mub[\psi(x)]}{\delta  \psi_{\alpha}(x)}
    \frac{\delta \mub^\dag [\psi(x)]}{\delta  \psi_{\beta}(x')}
 \ea
is the generalized Fisher matrix, and
 \be
    \label{llq}
    \frac{\delta \calL_0}{\delta \psi(x')} = -2 \Delta \!\D \, \C^{-1}
    \frac{\delta \mub^\dag [\psi(x)]}{\delta  \psi_{\beta}(x')}
 \ee
is the gradient of the log-likelihood. The maximum-likelihood
solution for the function is
 \be
    \delta \psi_\alpha(x) = \Delta \! \D \, \C^{-1} \int dx'
     \frac{\delta \mub^\dag [\psi(x)]}{\delta  \psi_{\beta}(x')}
     F^{-1}_{\alpha \beta}(x,x'),
 \ee
where again Greek indices represent a discrete infinite set of functions. 

\subsubsection{Flat prior}

We can now write the full path-integral marginalized
log-likelihood for a flat-prior as
 \be
 \label{mmml}
    \calL_0 = \Delta \D \C_M^{-1} \Delta \D^t + \Tr \ln \C_M,
 \ee
where
 \be
 \label{lll11}
    \C_M = [\C^{-1} - \C^{-1}\Pp\C^{-1}]^{-1}
 \ee
is the new marginalized covariance matrix,
 \be
    \Pp=\int \! dx' dx'' \,
    \frac{\delta \mub [\psi(x)]}{\delta  \psi_{\alpha}(x')}
    F_{\alpha\beta}^{-1}(x',x'')
    \frac{\delta \mub^\dag [\psi(x)]}{\delta  \psi_{\beta}(x'')},
    \label{flatprior}
 \ee
and we assume the functional Fisher matrix, $F_{\alpha
\beta}(x,x')$, is invertible. The marginalized covariance matrix,
$\C_M=\C_M(\thetab)$, is now a function of cosmological parameters
through the inclusion of the functional derivatives of the mean,
$\mub(\thetab)$.

\subsubsection{Gaussian prior}

In the presence of a Gaussian-prior, the form of the marginalized
likelihood and covariance matrix still holds, with the matrix
$\Pb$ now given by
 \be
    \Pp=\int \! dx' dx'' \,
    \frac{\delta \mub [\psi(x)]}{\delta  \psi_{\alpha}(x')}
    M_{\alpha\beta}^{-1}(x',x'')
    \frac{\delta \mub^\dag [\psi(x)]}{\delta  \psi_{\beta}(x'')}.
 \ee
Here the $M_{\alpha\beta}$ is a sum of the
path-integral Fisher and a prior covariance term
 \be
    M_{\alpha\beta}(x,x')=F_{\alpha\beta}(x,x')+C_{\alpha\beta}^{-1}(x,x').
 \ee
This contains information on the nuisance parameters from the
prior $C_{\alpha\beta}(x,x')$ and the data itself
$F_{\alpha\beta}(x,x')$. If we include this prior covariance on
the nuisance functions we find (see Appendix A for details) that
the marginalized likelihood is still a Gaussian in the data, with
a new covariance matrix;
 \be
 \label{csys}
    \C_M = \C + \int dx' dx'' \,
   C_{\alpha\beta}(x',x'') \frac{\delta
    \mub[\psi(x)]}{\delta
  \psi_\alpha(x')} \frac{\delta \mub^t[\psi(x)]}{\delta
  \psi_\beta(x'')}.
  \label{margcov}
 \ee
This form of the marginalized data covariance matrix has the clear
advantage that it does not require the inversion of a functional
Fisher matrix, or prior covariance matrix.

\subsection{Marginalized Fisher Matrices}

The Fisher Matrix for the marginalized likelihood is given by
(Tegmark, Taylor \& Heavens, 1997)
 \be
 \label{margF0}
    F_{ab} = \half \Tr [\mub_a  \mub_b^t+\mub_a \mub_b^t]\C_M^{-1},
 \ee
where Roman indices, $(a,b)$, denote cosmological parameters, and
for path-integral marginalized covariance $\C_M$ is given by equation 
(\ref{lll11}) for a flat prior and equation
(\ref{csys}) for a Gaussian prior. 
We have implicitly taken account of the cosmological
parameter information contained in the mean of the data via
$\mub(\thetab)$. However, even though the marginalized covariance
matrix, $\C_M$, now depends on cosmological parameter through the
functional derivatives of the mean, $\mub$, at the level of the
Fisher matrix $\C_M$ is a constant in parameter-space.

Throughout Section 2 we have made the assumption of a 
Gaussian-distributed function-space. If this assumption is 
dropped then the functional integration can be done numerically, 
one way to do this integration is using Monte-Carlo integration
techniques, or functional form-filling (Kitching et al., 2009). We
discuss these alternatives in Appendix B.

Having developed the formal aspects of path-integral
marginalization to cosmological likelihoods and Fisher matrices, we
now apply it to the case of marginalization over two cases of
systematic effect in Cosmology. The first is
scale-dependent galaxy bias, and the second is the case of
intrinsic alignments in weak gravitational lensing.

\section{Scale-dependent galaxy bias}
\label{Scale-dependent, stochastic galaxy bias}

As an example let us consider the case when the mean signal is an
estimate of the galaxy power spectrum measured from a galaxy
redshift survey;
 \be
    \mu(k,r) = P_{gg}(k)+N(r) ,
 \ee
where $N(r)=1/\bar n(r)$ is the distance-dependent galaxy
shot-noise and $\bar n(r)$ the mean galaxy number density.
The galaxy power spectrum is a biased representation of the underlying
matter power spectrum
 \be
    P_{gg}(k) = b^2(k) P_{\delta\delta}(k),
 \ee
where $P_{\delta\delta}(k)$ is the matter power spectrum and $b(k)$ is a
dimensionless, scale-dependent galaxy bias function. This
scale-dependent galaxy bias only depends on scale and not phase,
so is a function of $k=|\kk|$ only. While the bias function is, in
this model, deterministic, it is in general completely degenerate
with the matter power spectrum. Hence we need to introduce a
Gaussian prior from the start, where the prior knowledge could
come from weak lensing measurements (e.g. van Waerbeke, 1998; Bernstein, 2009) or a
semi-analytic galaxy bias model.
In either case there will be some
level of uncertainty associated with the bias model, $\delta
\psi(k) = \delta b(k)$. We can, of course, extend this model to
include stochastic bias effects (e.g., Dekel \& Lahav, 1999).

In the following we include the effect of linear redshift-space distortions by the
transformation (Kaiser, 1987)
 \be
    P^s_{gg}(\kk)=[b(k)+f \mu_k^2]^2 P_{\delta\delta}(k),
 \ee
where $f=d\ln \delta_m/d\ln a$ is the growth rate of matter
perturbations, and $\mu_k$ is the cosine of the angle between the
wavevector $\kk$ and the line-of-sight. We shall assume we are
working on scales large enough to ignore non-linear finger-of-god
effects. If we assume the power is a Gaussian random variable, the
covariance of the galaxy power is given by
 \be
    C_k = 2 \left[ P^s_{gg}(\kk)+N(r)\right]^2.
 \ee
Applying
equation (\ref{csys}), assuming a Gaussian prior, 
the functionally marginalized covariance is
 \be
    C_M(\kk)=  2 [P_{gg}^s(\kk)+N(r)]^2+ B(\kk),
 \ee
where
 \be
    B(\kk) = 4 \left(\frac{P_b(k)}{b^2(k)}\right)|P^s_{gg}(\kk)|^2
 \ee
is the increase in the covariance matrix after we have
marginalized over all bias functions. 
As the galaxy bias is independent of phase, and  both the galaxy
and matter distributions are statistically homogeneous, we only
need to know the power, or functional variance, in the bias;
\be
P_b(k) = \lgl |\delta b(k)|^2\rgl,
\ee
were the averaging is taken over all functions of $b(k)$.

If we substitute the redshifted galaxy power spectrum  into the
Fisher Matrix formalism, (see e.g., Taylor \& Watts, 2001; Seo \&
Eisenstein, 2003; Burkey \& Taylor, 2004), we find the Fisher
Matrix including path-integral marginalization over all possible
bias functions is
 \ba
    F_{ab} &=& \half \int \!\frac{d^3 k}{(2 \pi)^3}
    \frac{\de \ln P^s_{gg}(\kk)}{\de \theta_a}
    \frac{\de \ln P_{gg}^s(\kk)}{\de \theta_b} V_{\rm eff}(\kk),
 \ea
where
 \be
    \label{vvv}
    V_{\rm eff}(\kk) = \int \! d^3r
    \left(\frac{[P_{gg}^s(\kk)]^2}{2[P^s_{gg}(\kk)+N(r)]^2+B(\kk)}\right),
 \ee
is the effective volume of the survey. 
As $B(k)$ also depends on
$P_{gg}^2(k)$, the effective volume fully specifies our survey. 
In the limit of no
systematic uncertainty in the bias, $P_b(k)=0$, or $B(k)=0$, this
reduces to the usual Fisher matrix for the redshifted power
spectrum.

\begin{figure}
  \centerline{\includegraphics[width=1.15\columnwidth,angle=0,clip=]{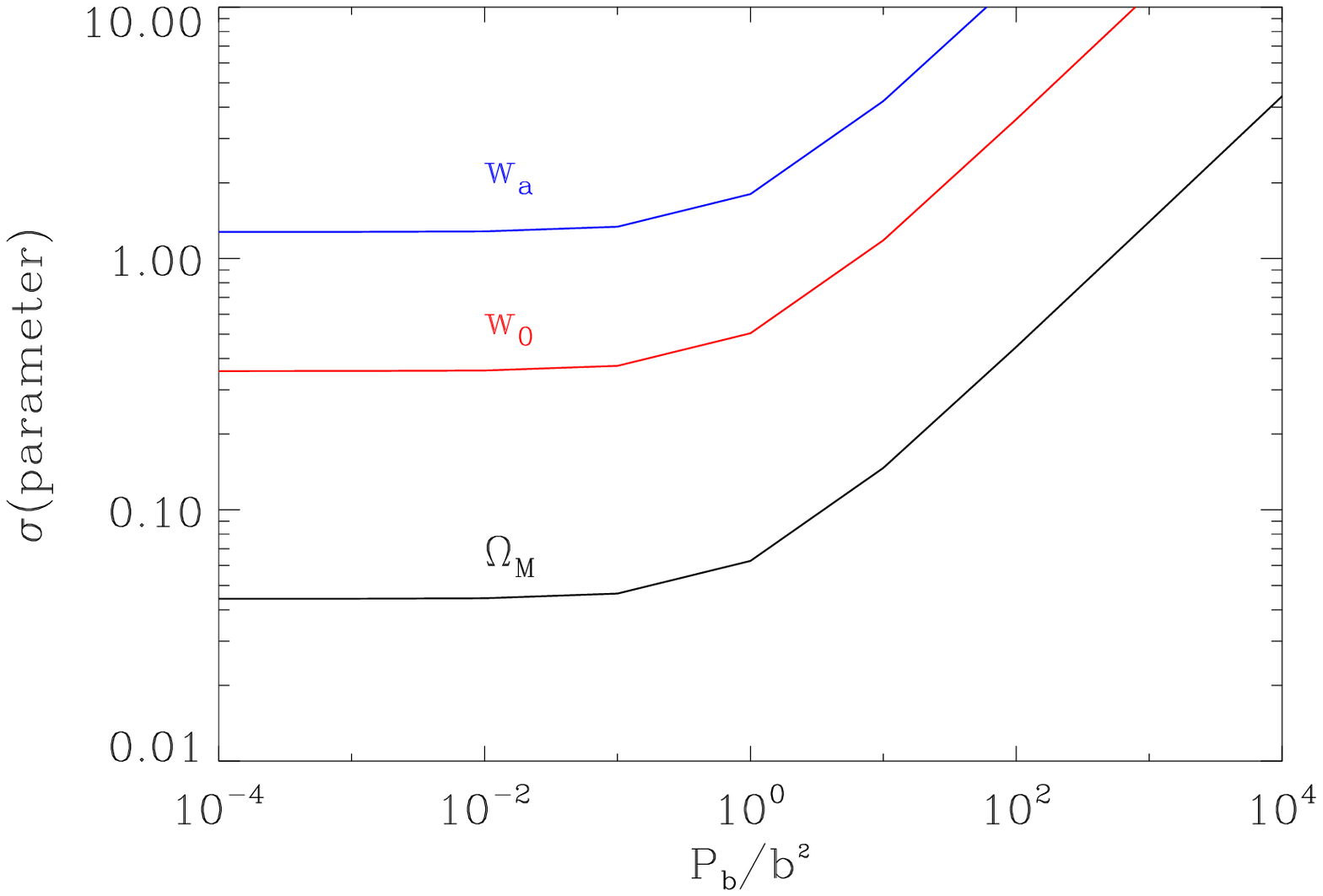}}
 \caption{The marginalized errors on $\Omega_m$ (lower, black
   line), $w_0$ (middle, red line)
   and $w_a$ (upper, blue line) for a Euclid
   spectroscopic survey (Laureijs et al., 2009) marginalized over all
   possible bias functions $b(k)$. We varie the functional variance 
   $P_b(k)/b^2(k)$ which weights the space of functions that
   are marginalized over about a fiducial function. In the
   limit that $P_b(k)/b^2(k) \rightarrow\infty$ all possible functions are marginalized
   over, in the limit that $P_b(k)/b^2(k) \rightarrow 0$ the bias is fixed at the
   fiducial function.}
 \label{pbias}
\end{figure}

We demonstrate the effect of path-integration over all possible
galaxy scale-dependent bias functions with a Gaussian prior in 
Figure \ref{pbias},
where we consider a Euclid spectroscopic survey (Laureijs et al.,
2009) with a survey volume of $19.7$ h$^{-3}$Gpc$^3$ and a maximum
wavenumber of $0.5 h {\rm Mpc}^{-1}$. We shall assume that the
signal-to-noise ratio of the survey is unity, $\bar n P_{gg}=1$ (Seo \&
Eisenstein, 2003). We use a cosmological parameter set that
allows for curved cosmologies with parameters $\Omega_m$,
$\Omega_{de}$, $\Omega_b$, $h$, $\sigma_8$, and $n_s$ given by
$(0.25,0.7,0.75,0.8,0.95)$, and parameterize the dark energy
equation-of-state using a first-order Taylor expansion,
$w(z)=w_0+(1-a)w_a$ (Linder, 2003; Chevallier and Polarski, 2001)
with $(w_0,w_a)=(-0.95, 0.0)$. We assume a fiducial bias
function of $b(k)=1$. In Figure \ref{pbias} we show the
marginal errors on $\Omega_m$, $w_0$ and $w_a$ respectively, as a
function of the functional variance in the bias. In the limit that
the bias variance is small the accuracy converges towards the
systematic-free case. However, as the bias variance increases
beyond $P_b(k)/b^2(k) \approx 10^{-1}$, the uncertainty on 
cosmological parameters grows large. Using this result and from
equation (\ref{vvv}) we define an approximate scaling relation 
\be 
\frac{P_b(k)}{b^2(k)}\ls 0.22\left(1+\frac{1}{\bar n P_{gg}}\right), 
\ee
for cosmological parameter constraints to be unaffected by uncertainty
in the scale-dependent bias. 

Constraints on the bias function of $P_b(k)/b^2(k) \approx 10^{-1}$ 
can be achieved using large area surveys and techniques such as shear-galaxy cross
correlations (e.g. van Waerbeke, 1998; Bernstein, 2009) or
bispectrum statistics (e.g. Matarrese et al., 1997; Verde et al.,
1998; Guo \& Jing, 2009). A more complete analysis should include
the redshift-dependence of galaxy bias, as well as
scale-dependence, and also consider stochastic bias effects.
However, consideration of equation (27) suggests that all of these effects
will only become important if they become larger than the sampling
and shot noise terms.

\section{Cosmic Shear Intrinsic Alignments}
\label{Cosmic Shear Intrinsic Alignments}

We now consider the effect of intrinsic alignments (IA) on weak
lensing. Cosmic shear uses the weak lensing distortion caused by
large-scale structure along the line of sight as a probe of
cosmology; this assumes that galaxies are randomly aligned, so
that the mean observed ellipticity is zero. However there are two
effects which act to align galaxies by adding `intrinsic' ellipticity
which is not part of the cosmic shear signal. Intrinsic-Intrinsic
(II) alignment is the small-scale effect of galaxies aligning due,
for example, to local tidal forces. We will ignore this effect in
our example since this can be removed by neglecting close galaxy
pairs in angle and redshift from the cosmic shear signal (e.g
Heymans \& Heavens, 2003). The Galaxy, or shear, Intrinsic (GI)
alignment (Hirata \& Seljak, 2004) is the correlation between the
foreground galaxy ellipticity and the induced shear in a
background galaxy and this is more difficult to correct for.

The tomographic cosmic shear power spectra $C^{GG}_{ij}(\ell)$, 
where the two-point correlation of the shear field is binned in
redshift, can be written as
 \be
 \label{GG}
    C^{GG}_{ij}(\ell)=\int_0^{r_H}\! dr
    \, W^{\rm GG}_{ij}(r)
    P_{\delta\delta}\!\left(\frac{\ell}{S_k(r)};r \right),
 \ee
where lensing weight can be expressed as
 \be
 W^{\rm GG}_{ij}(r)=\frac{q_i(r)q_j(r)}{S^2_k(r)},
\ee
and the kernel is 
\be
q_i(r) = \frac{3 H^2_0 \Omega_m S_k(r)}{2 a(r)}
\int_r^{r_H}\!  dr' \, p_i(r')
\frac{S_k(r'-r)}{S_k(r')}.
\ee
We follow the notation of Joachimi \& Bridle (2009).  The
comoving distance is $r$, $r_H$ is the horizon distance, while
$S_k(r)=\sin(r), r,
\sinh(r)$ for curvatures $k=-1,0,+1$, $a$ is the scale factor
and $P_{\delta\delta}(k;r)$ is the 3D density-density 
matter power spectrum. The comoving galaxy probability 
distribution is given by $p_i(r)$. The
$ij$ subscripts refer to redshift bins, where the shear field is
approximated as a series of correlated 2D planes. This can be
generalized to the full 3D shear field (Kitching \& Heavens,
2010).

In a similar way the GI power spectrum can be written as an
integral over the density-shear power spectrum;
 \ba
    \label{GGI}
    C^{GI}_{ij}(\ell)=\int_0^{r_H} \! dr \,
    W^{GI}_{ij}(r)
    P_{\delta\gamma}\!\left(\frac{\ell}{S_k(r)};r \right),
 \ea
where we have defined the GI weight function as
 \be
    W^{GI}_{ij}(r)=n_i(r)q_j(r)/S^2_k(r),
 \ee
and $n_i(z)$ is the number density of galaxies in the redshift bin
$i$. We can now write the mean signal as
 \be
\label{lensmean}
    \mu_{\nu}(\ell)=C^{GG}_{ij}(\ell)+C^{GI}_{ij,j>i}(\ell)+N_{ij},
 \ee
where Greek subscripts $\nu=(i,j)$ denote pairs of redshift bins, and
the notation $j>i$ indicates that the matrix is non-zero when the
lensed galaxy is behind the intrinsically aligned galaxy. The
shear shot noise term $N_{ij}=\delta_{ij}\sigma^2_{\epsilon}/2\bar
n_i$ (Hu, 1999), where $\sigma_{\epsilon}$ is the variance in the
intrinsic ellipticity distribution, only adds to the $C^{GG}_{ij}$
term. The covariance of the mean is
 \ba
    C_{\mu\nu}(\ell)=
    C^{GG}_{jm}(\ell) \left[ C^{GG}_{il}(\ell) + C^{GI}_{il,i>l}(\ell)
    + C^{GI}_{il,i<l}(\ell)\right] \nn
   +
    \left[C^{GG}_{im}(\ell)+C^{GI}_{im,i>m}(\ell) \right]
    \left[C^{GG}_{jl}(\ell)+C^{GI}_{jl,l>j}(\ell)\right].
 \ea

Currently there are two approaches to removing the GI contaminant;
nulling and modelling. In nulling the GI contaminant is removed
from the data by transforming the data vector, by ordering the
transformed modes in order of contamination, and removing the most
contaminated ones (Joachimi \& Schnieder, 2008, 2009). However,
nulling also removes cosmological information and can bias the
recovered cosmological parameters.

The second approach is modelling, which uses a phenomenological
(e.g. Heymans \& Heavens, 2003) or theoretically motivated model
(e.g. Schneider \& Bridle, 2009) and  marginalizes over the parameters
in that model. The problem with the modelling approach is that it is 
sensitive to the choice of model and the resulting cosmological
parameter errors are dependent on this choice (Bridle \& King,
2007; Kitching et al., 2008; Joachimi \& Bridle, 2009).
Furthermore numerically marginalizing over $100$'s of parameters,
required to characterize the model, is computationally expensive,
although semi-analytic marginalization could be used to reduce
this expense (Taylor \& Kitching, 2010).

An alternative, third way, which we propose here is to marginalize
over all possible intrinsic alignment functions using path-integral
marginalization.

\subsection{ Intrinsic Alignment Removal}

In this Section we investigate two distinct cases. The first case
is marginalization over the uncertainty in the density-ellipticity
relation itself. In this first case we assume both a flat-prior
(the self-calibration case), and a Gaussian-prior. A Gaussian prior
may use information on the density-ellipticity relation gained from, for
example, galaxy-galaxy lensing. In the second case we make some
assumptions about how the density-ellipticity power spectrum is
related to the matter power spectrum via a set of
intrinsic-alignment bias functions, and assume only a Gaussian
prior.

\subsubsection{Intrinsic Alignment removal: flat prior}
\label{Marginalizing over Intrinsic Alignment amplitude}

The functional variation of the IA signal with the
density-ellipticity cross-spectrum is given by (see Appendix C) 
 \be
 \frac{\delta C^{GI}_{ij,j>i}(\ell)}{\delta P_{\gamma\delta}(\ell'/r,r)}
  =W^{GI}_{ij,j>i}(r) \frac{ 2 \pi \delta_D(\ell-\ell')}{\ell}.
 \ee
For a flat-prior, where we use the data itself to fit and
marginalize over the IA term, we use equations (\ref{lll11}) and
(\ref{flatprior}) to find the marginalized covariance
 \ba
 \label{ffgghe}
 \C^M_{\mu\nu}(\ell) = [\C^{-1}_{\mu\nu}(\ell) -
   \C_{\mu\mu'}^{-1}(\ell)\Pp_{\mu'\nu'}(\ell)\C^{-1}_{\nu'\nu}(\ell)]^{-1}
\ea
where
\ba
 \Pb_{\mu\nu}(\ell) = \int_0^{r_H} \!\! dr dr' \,
    W^{GI}_{\mu}(r)[F(\ell; r,
    r')]^{-1} W^{GI}_{\nu}(r'),
 \ea
and the functional Fisher matrix for the IA power spectrum is
\be
\label{ffgghf}
F(\ell;r,r')
=\sum_{\mu\nu}\C^{-1}_{\mu\nu}(\ell)W^{GI}_{\mu}(r)W^{GI}_{\nu}(r').
\ee
To invert the functional Fisher matrix we bin the radial GI
weights in equations (\ref{ffgghf}) and (\ref{ffgghe}) using $200$
radial bins, while the angular terms are always diagonal.
The Fisher matrix for cosmological parameters is given 
by equation (\ref{margF0}), this is more explicitly shown for the IA
case in Appendix C. 

To make a qualitative assessment of the impact of intrinsic
alignment removal we consider a Euclid 
weak lensing survey (Refregier et al., 2010) 
of $20,000$ square degree, with median
redshift of $z_m=1.0$ and $35$ galaxies per square arcminute with
a photometric redshift uncertainty of $0.03(1+z)$. We assume the
galaxy number density is given by  $n(z)\propto z^2{\rm
exp}(-1.4z/z_m)^{1.5}$, and use the same cosmological
model as in Section 3.

To quantify the effect of path-integral IA removal on dark energy
surveys we use the dark energy Figure-of-Merit (FoM; Albrecht et
al., 2006), defined as the area constrained in the uncorrelated 
($w_0$,$w_a$) plane with
${\rm FoM}=1/[\sqrt(F^{-1}_{w0w0}F^{-1}_{wawa}-(F^{-1}_{w0wa})^2)]$. We
assume the fiducial model for the GI Intrinsic Alignment is given
by the linear/nonlinear alignment model of Bridle \& King (2007;
see also Section \ref{The Intrinsic Alignment Model})\footnote{Our
code is an extension of the {\tt iCosmo} package (Refregier et
al., 2008) and is available on request}.

For comparison we find that, for a default model with fixed IA
and no cosmological dependence of the IA term (i.e.,
cosmic shear only, treating IA as additional known noise term), 
the ${\rm FoM}=130$. If we allow the
ellipticity-density relation to be fixed by the
data, with a flat-prior and allow a cosmological dependence in the
weak-lensing weight of the GI term, we find the FoM is degraded to
${\rm FoM}\approx 100$ (see Figure \ref{vbias_pd}). 
In this case, information is gained because of the extra
cosmological dependence of the GI weights, but also lost as the data
now has to measure the density-ellipticity relation.

It is interesting to compare the path-integral IA removal approach
to the nulling method of Joachimi \& Schneider  (2008, 2009) which
uses the radial information in the lensing kernel to null the GI
term from the shear data vector. Nulling makes a number of
assumptions about the lens and source planes being thin sheets,
and fixes the cosmology in the lensing kernel. An effect of nulling is
to introduce a bias
in cosmological parameters due to its effect on the shear power.
Nulling finds a factor $\sim 2$ reduction
in a global FoM, combining errors from all cosmological
parameters, which seems comparable to path-integral
marginalization. We leave a full comparison of methods for future
work.

\subsubsection{Intrinsic Alignment removal: Gaussian prior}

If, instead of self-calibration, we use external data to
constrain IA with a Gaussian-prior, we find the path-integral
marginalized covariance, equation (\ref{margcov}), can be written
as (see Appendix C) 
\ba
\label{margC2a}
\C^M_{\mu\nu}(\ell)=\C_{\mu\nu}(\ell)+
\int_0^{r_H} \!\! dr \,  \sigma^2_P(\ell/r,r) W^{GI}_{ij}(r)
W^{GI}_{lm} (r),
\ea
where $\sigma^2_P(\ell/r,r)=\lgl |\delta
P_{\gamma\delta}(\ell/r,r)|^2\rgl$ is the prior uncertainty in the
shear-density relation.  We again make no assumption about the
relationship between matter and ellipticity. The path-integral
Fisher matrix, marginalizing over all possible shear-density
cross-power, is again given by equation (\ref{margF0}) and in Appendix
C. 

\begin{figure}
  \centerline{\includegraphics[angle=0,clip=,width=1.\columnwidth]{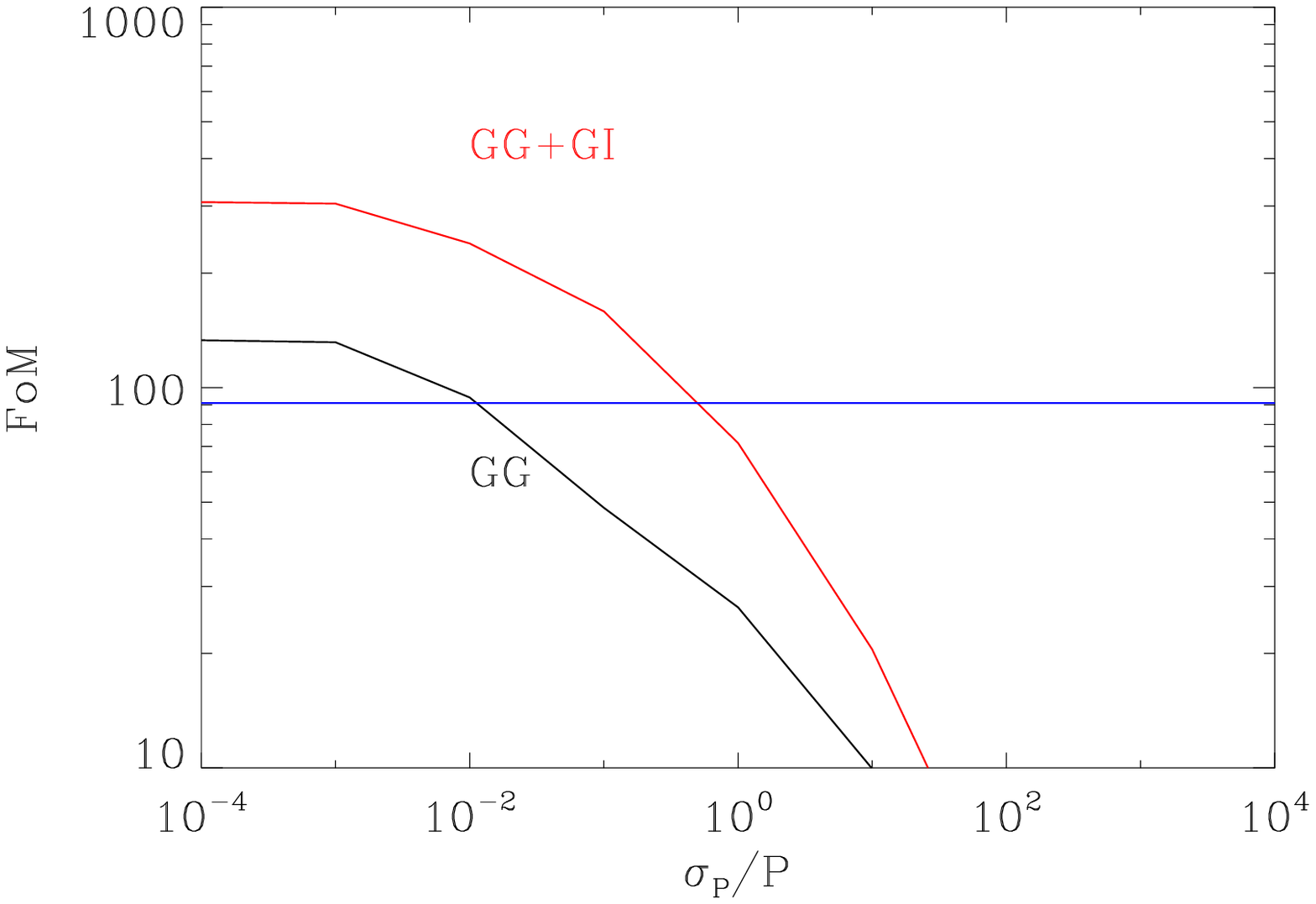}}
 \caption{The dark energy Figure of Merit (FoM) for a Euclid-like
   tomographic weak lensing survey (Refregier et al., 2010)
   marginalized over all possible non-linear alignment bias functions.
   We show results as a function of the fractional scatter in the
   $P_{\delta\gamma}(k)$ function-space, assuming a Gaussian prior.
   The lower black line (GG) shows the FoM from cosmic shear alone, 
   assuming the GI term does not depend on
   cosmological parameters.
   The upper red line (GG$+$GI) shows the FoM when we also allow the GI 
   redshift dependence to vary with cosmological parameters.
   The horizontal (blue) line shows the FoM for the self-calibration
   case in which flat prior in function-space is assumed (equation
   \ref{ffgghe}) -- in this
   case the shear and GI power spectra are dependent on cosmology
   (similar to the upper red line).}
 \label{vbias_pd}
\end{figure}

In Figure \ref{vbias_pd} we show how the dark energy FoM changes
as the fractional scatter in the ellipticity-density power
spectrum;  $\sigma[P_{\gamma\delta}(\ell/r,r)]/P_{\gamma
\delta}(\ell/r,r)$. The lower solid line extends our default model
from Section \ref{Marginalizing over Intrinsic Alignment
amplitude}, by treating the IA term as additional noise where 
the GI term is independent of cosmological parameters but
the ellipticity-shear relation is allowed to vary (constrained by
the prior on $P_{\gamma \delta}(\ell/r,r)$). 
In this case the FoM is rapidly
damped from ${\rm FoM}=130$, as the
scatter is increased above $\sigma(P_{\gamma
\delta})/P_{\gamma \delta} \approx 10^{-3}$. As anticipated in
Section 2.2, as the scatter is allowed to increase this adds
greater uncertainty to the model, with ever larger fluctuations in
the ellipticity-density relation. As a result the information is
actually damped below that of the flat-prior.

The upper curve in Figure \ref{vbias_pd} allows the 
redshift-dependence of the GI weights to vary with cosmology and act
as a source of information. This increases the FoM, in the limit
of no uncertainty in the IA term, to ${\rm FoM}=300$, showing that the
known lensing dependence of
the GI term contains useful cosmological information. Even though 
the FoM drops off rapidly, as the scatter is increased, it still
remains a factor of a few greater than ignoring the cosmological
information in the GI term.

\begin{figure*}
  \includegraphics[angle=0,clip=,width=\columnwidth]{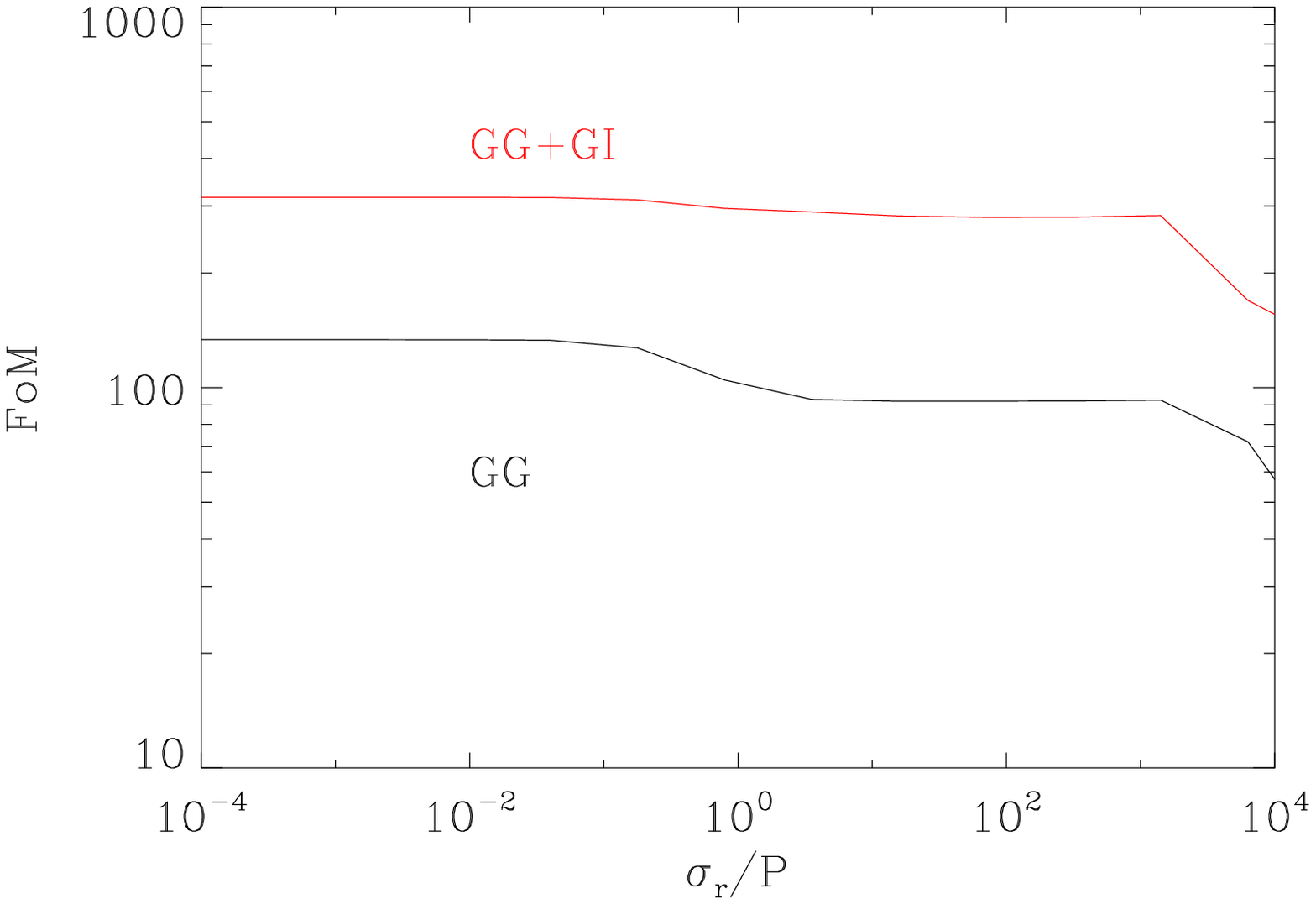}
  \includegraphics[angle=0,clip=,width=\columnwidth]{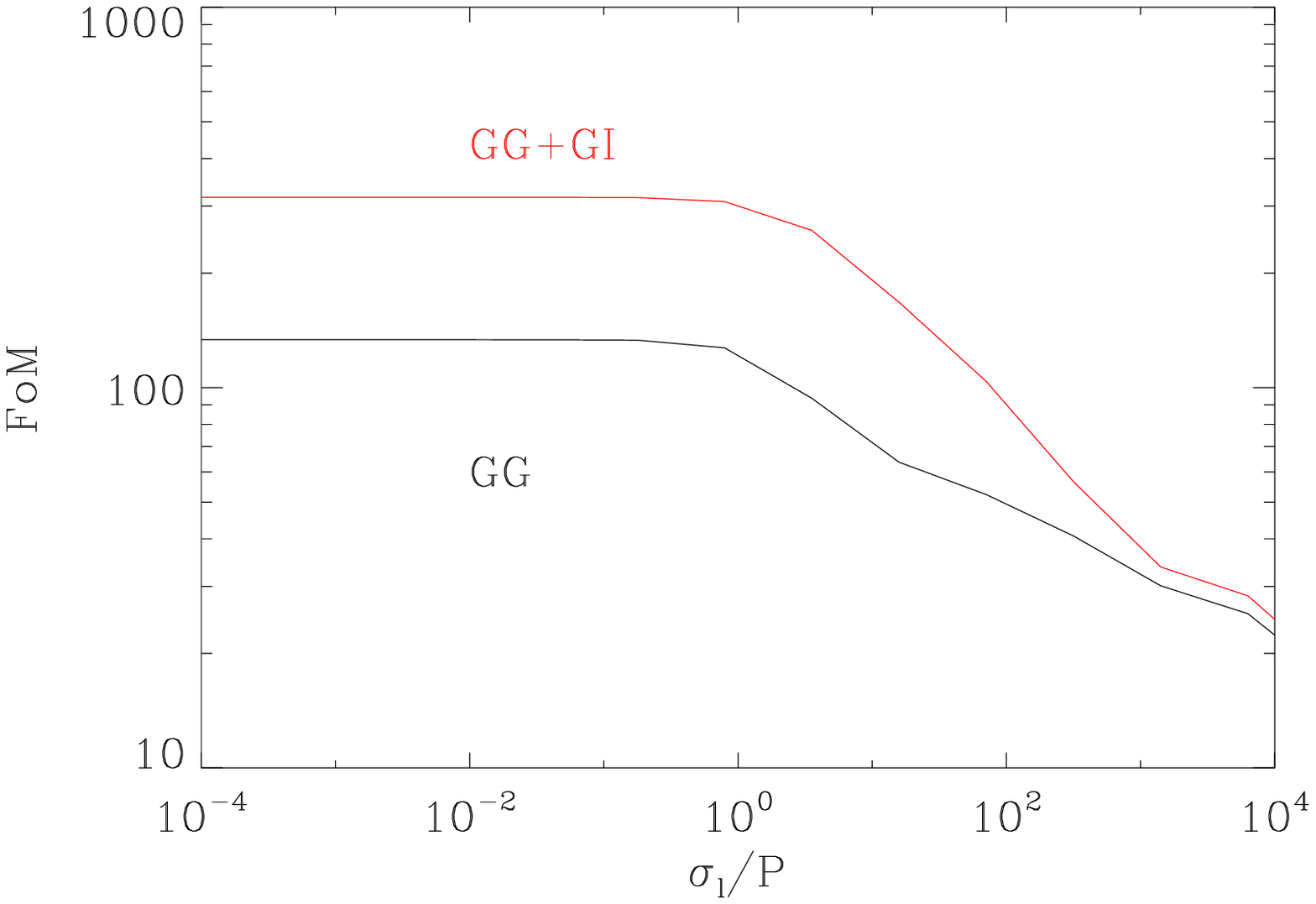}
 \caption{The dark energy Figure of Merit (FoM) for a Euclid-like
   tomographic weak lensing survey
   marginalized over all non-linear intrinsic alignment bias functions.
   We show results as a function of the scatter in radial intrinsic
   alignment bias,
   $b_{I}(r)$ with the angular bias fixed at $\sigma_b(\ell)/P=10^{-4}$
   (left panel) and angular intrinsic
   alignment bias, $b^I(\ell)$ with
   $\sigma_b(r)/P=10^{-4}$ (right panel) (effectively tracing the
   along the x and y-axes in Figure \ref{vbias2D}).
   The lower (black) lines show the FoM when only the shear signal has a cosmological
   dependence and the GI term is treated as noise.
   The upper (red) lines show the FoM when both the shear
   and the radial weak lensing dependence in the GI term dependent
   on cosmology.}
 \label{vbias}
\end{figure*}

\begin{figure}
  \centerline{\includegraphics[angle=0,clip=,width=1.1\columnwidth]{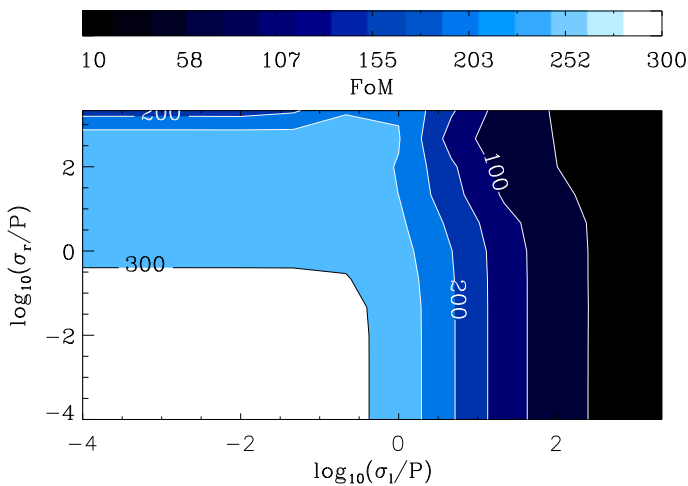}}
 \caption{Dark energy Figure of Merit (FoM) contours, for a Euclid-like
   tomographic weak lensing survey
   marginalized over all possible non-linear alignment bias functions,
   as a function of the fractional scatter in radial intrinsic
   alignment bias (y-axis) and angular intrinsic alignment bias (x-axis).
   Here we have assumed the GI term is dependent on cosmology (upper
   lines in Figure \ref{vbias}).
   The FoM reduction is more severe for a given uncertainty in the
   angular intrinsic alignment bias.}
 \label{vbias2D}
\end{figure}

\subsubsection{The Intrinsic Alignment Model}
\label{The Intrinsic Alignment Model}

In this Section we will assume that the shear-density power
spectrum, $P_{\gamma\delta}$, is related to the matter power
spectrum by some unknown ellipticity bias which is a function of
redshift and scale (e.g., Bernstein, 2009)
 \be
    \label{nla}
    P_{\gamma\delta}(\ell/r,r)
    =b_I(\ell)b_I(r)P_{\delta\delta}(\ell/r,r),
 \ee
and that the dimensionless bias function is separable in radial
distance $r$ and
angular wavenumber $\ell$. In this case the GI effect can be written in the same way
as the tomographic shear power spectrum;
\be
\label{GI}
C^{GI}_{ij}(\ell)=\int_0^{r_H} \!\!\! d r \,
W^{\rm GI}_{ij}(r)
b_I(\ell)b_I(r)P_{\delta\delta}\!\left(\frac{\ell}{S_k(r)};r\right).
\ee
This is similar to the (linear/non-linear) alignment model
introduced in Hirata \& Seljak (2004) and generalized by Bridle \&
King (2007) and Joachimi \& Bridle (2009). The intrinsic alignment
bias relation, equation (\ref{nla}), assumes that the intrinsic
alignment is a non-local convolution of the angular density field
which depends locally on distance.

Given we are assuming a model for which we should have some
motivation to trust, we shall only assume a Gaussian-prior on the
IA bias functions. In Appendix D we show that the path-integral
marginalized covariance for this model is given by
\be
\label{margC0}
C^M_{\mu\nu}(\ell) = C_{\mu\nu}(\ell)+ B_{\mu\nu}(\ell)
\ee
where
\ba
B_{\mu\nu}(\ell) \!\!\!\!\!&=&\!\!\!\!\!
b_I^2(\ell) \int_0^{r_H} \!\! d r \, \sigma^2_b(r)
W^{GI}_{ij}(r) W^{GI}_{lm}(r) \left|P_{\delta\delta}(\ell/r;r)\right|^2
\nn
& &+
\left( \frac{\sigma_b^2(\ell)}{b_I^2(\ell)}\right)
C^{GI}_{ij}(\ell) C^{GI}_{lm}(\ell).
\ea
We assume $b_I(r[z])=C_1\bar\rho(z)/D(z)(1+z)$ and $b_I(\ell)=1$
as the fiducial functions for the biases (see Bridle \& King, 2007
and Joachimi \& Bridle, 2009 for these definitions) and that the
scatter in these functions is given by $\sigma_b(r)$ and
$\sigma_b(\ell)$. Equation (\ref{margF0}) is again used to derive
a Fisher matrix for the non-linear alignment bias model (also given in
Appendix C). 

Figure \ref{vbias} shows the dark energy FoM as a function of the
radial (left plot) and angular scatter (right plot) in the IA bias,
for our fiducial survey. 
The lower set of lines assumes that the GI terms are independent
of cosmology, while the upper set of lines includes the
cosmological dependence of the GI terms. We see that including
the GI cosmology dependence increases the FoM by a factor
of around three. 
The angular bias reduces the FoM when $\sigma_b(\ell)/P \gs
10^{-1}$, while the radial bias does not affect the FoM until
$\sigma_b(r)/P\gs 10^3$. This is because the
dominant effect of the dark energy equation of state is to change
the angular part of the lensing power spectrum. While the growth
rate is also effected by changes in the dark energy equation of
state, the extra geometric constraints from the lensing kernel (in
both the shear and GI terms) help to break any degeneracies
introduced. As a result the dependence of the FoM on the radial
functional scatter is weak (and even more so when both shear and
GI contribute to the geometric constraints). 
This is also in
agreement with Joachimi \& Bridle (2009) where a large but
finite-dimensional marginalization has found a similar behaviour.

Figure \ref{vbias2D} shows a 2-D contour plot of the dark energy
Figure of Merit as a function of the fractional scatter in radial
and angular intrinsic
alignment bias. Again it is clear that marginalizing over angular,
$\ell$-dependent 
bias has a larger impact than the radial bias on the FoM. 
A constraint in the functional scatter of $\sigma_b/P\sim 10^{-1}$,
required such that the FoM is unaffected, is achievable using
current techniques applied to future surveys, 
e.g. Mandelbaum et al., 2009; who find a GI functional constraint of 
$\sim 1$ using WiggleZ.

\section{Conclusions}
\label{Conclusions}

Path-integral marginalization can account for the impact of measuring
and marginalizing over unknown functions on cosmological parameter
estimation. We have constructed a likelihood function in cosmological 
parameter space
that takes into account marginalization over unknown functions,
assuming either a flat-prior, were we use information in the data
itself to measure and remove the effect, or a Gaussian-prior,
where we use external data to constrain the nuisance functions.
We assume the distribution in
function-space is well approximated by a Gaussian functional, but note
that the non-Gaussian case can be treated with numerical methods.

We use the likelihood to derive an expression for the cosmological
parameter Fisher information matrix in the case that all functions
are marginalized over. We have applied this to the case of a
scale-dependent galaxy bias function, relating the galaxy power
spectrum to the underlying matter power spectrum. As a
scale-dependent bias is completely degenerate with the matter
power spectrum we assume a Gaussian-prior with variable scatter.
We show that the marginal errors of $\Omega_m$,
$w_0$ and $w_a$ rapidly increase as the uncertainty 
in the fractional power of 
a scale-dependent bias increases above $P_b(k)/b^2(k)\gs 10^{-1}$, 
implying we need to constrain any
scale-dependent bias to a few percent accuracy. We did not explore
redshift-dependent bias models here, but the marginalized Fisher
matrix implies a similar accuracy is needed. In addition we did
not explore stochastic bias models, which will add a further
requirement. There are clear ways to constrain scale-dependent bias 
from redshift-space distortions, which also probe the velocity field and so
are bias-free; by the galaxy bispectrum; or by combining galaxy
redshift surveys with weak lensing probes of the dark matter 
(e.g. van Waerbeke, 1998; Bernstein, 2009; 
Matarrese et al., 1997; Verde et al., 1998; Guo \& Jing, 2009).

Our second example is in the removal of Intrinsic Alignments
effects, in particular the non-local GI effect, on cosmic shear
measurements. In addition to nulling (Joachimi \& Schneider, 2008,
2009) and modelling (Bridle \& King, 2007; Kitching et al., 2008;
Joachimi \& Bridle, 2009), path-integral marginalization represents a third way to
remove and assess the impact of this effect. Path-integral
marginalization has the advantage that it is model-free, and
introduces no biases in cosmological parameter estimates. For a
flat-prior on the ellipticity-density relation, the dark energy
FoM drops from a fiducial ${\rm FoM}=130$, to ${\rm FoM}\approx 100$, 
when we allow the GI term
to be completely free in both angular and radial dependence and measured by
the data.

If we assume the GI ellipticity-density relation is accurately
constrained from external data (e.g. Mandelbaum et al., 2009) or
complimentary information (e.g. shear-position information;
Bernstein, 2009) with a Gaussian-prior, and allow the weak lensing
effect in the GI term to depend on cosmology, we find that the
dark energy constraint is enhanced from FoM$=130$ to FoM$=300$.
The FoM is degraded as the accuracy drops below $\sigma(P_{\gamma
\delta})/P_{\gamma\delta}\approx 10^{-3}$. 
The assumption of a non-linear alignment model (e.g
Hirata \& Seljak, 2004; Bridle \& King, 2007) relaxes this
requirement to $\approx 10^{-1}$.

Our results imply the GI effect in weak lensing can
be removed by path-integral marginalization and that cosmic shear
can still constrain dark energy to high precision. In addition, with no extra
assumptions, we have shown that the GI effect can 
itself be used to probe cosmology through its dependence on
weak lensing. By constraining the GI ellipticity-density
dependence, using external data, we can use the weak lensing 
dependence in GI to increase our knowledge of dark energy.

The path-integral marginalization approach presented here is a powerful
and general method to assess, control and remove the effects of
systematics which cannot be simply modelled as parametric
functions.


\vspace{0.5cm}

\noindent{\em Acknowledgements:} We thank Fergus Simpson, Adam
Amara, Benjamin Joachimi, Andrew Liddle, John
Peacock and Alan Heavens for useful discussions.
We also thank the DUEL network (MRTN-CT-2006-036133) for
supporting this work. TDK is supported by STFC rolling grant
number RA0888.


\onecolumn

\section*{Appendix A: Marginalization of systematics in the mean}

We start with the likelihood function where the cosmological
parameters, $\thetab$, and nuisance parameters, $\psib$, both
appear in the mean, $\mu(\thetab,\psib)$. Expanding the nuisance
parameter to first order and explicitly marginalizing over
variations in the nuisance parameters, $\psib$, we find,
\be
    L(\D|\thetab) = \int d [\delta \psi] \,
    \frac{e^{-1/2 (\Delta {\small \D}-\delta \psi_i \,\de_{\psi_i} \!\mub)\C^{-1}
    ({\small \Delta \D}-\delta \psi_j \,\de_{\psi_j} \!\mub)^t}
    }{\sqrt{(2 \pi)^N\det \C}} \frac{e^{-\delta {\small \psib}
    \C_{\psi}^{-1} \delta {\small \psib^t} /2}}{\sqrt{(2 \pi)^N_\psi\det \C_{\psi}}}.
\ee
First we expand in a Fourier series
 \ba
    f(\x,\mub,\C) &=&
    \int d^m s  \int d^n k \, e^{-\half \kk \C \kk^t} e^{-i
    \kk.(\x- s_i \de_i \mub)} \int d^m k' \, e^{-\half \kk' \C_s \kk'^t } e^{-i
    \kk'. \s}
    \nn
    &=& \int d^n k \, e^{-\half \kk \C \kk^t} e^{-i
    \kk.\x} \int d^m k' \, e^{-\half \kk' \C_s \kk'^t} \int d^ms \, e^{-i
    (\de_i \mub. \kk-k_i')s_i} \nn
    &=& \int d^n k \, e^{-\half \kk \C \kk^t}
     e^{\half (\kk. \de_i \mub)  C_{s,ij}(\de_j \mub. \kk)    }
       e^{-i \kk.\x} \nn
    &=&\frac{e^{-\half \x [\C+  (\de_i \mub)  C_{ij}^s (\de_j\mub^t)]^{-1} \x^t}}{
    \sqrt{(2 \pi)^n\det [\C+(\de_i \mub)  C_{ij}^s (\de_j\mub^t)]}}
 \ea
If we carry out the Gaussian integration this reduces to
 \be
    \label{ml} L(\D|\theta) = ((2 \pi)^N \det[\C+ C^{\psi}_{ij}
    (\de_{\psi_i} \! \mub)(\de_{\psi_j} \! \mub^t)])^{-1/2}\exp
    \left(-\half \Delta \D [\C+ C^{\psi}_{ij}
    (\de_{\psi_i} \! \mub)(\de_{\psi_j} \! \mub)^t]^{-1} \Delta \D^t \right).
 \ee
We note that this can be encapsulated by a modification of the
covariance
 \be \label{mc}
   \C_M =  \C+ C^{\psi}_{ij}
  (\de_{\psi_i} \! \mub)(\de_{\psi_j} \! \mub^t),
 \ee
where $C^{\psi}_{ij}$ is the covariance of the systematic, which
acts as an effective prior. This is a generalization of the
results from Bridle et al., (2002) and Taylor \& Kitching (2010)
where $\Ct = \C + \sigma^2_{\delta A} \mub \mub^t$. After carrying
out a path-integral marginalization over the bias functions, the
new data covariance matrix is given by
 \be
\label{lll0}
 \C_m = \C + \int dx dx' \, \lgl \psi_i(x) \psi_j(x') \rgl
    \frac{\delta \mub[\psi(x)]}{\delta
  \psi_i(x)} \frac{\delta \mub^t[\psi(x')]}{\delta
  \psi_j(x')},
 \ee
where the derivatives $\delta \mub[\psi(x)]/\delta\psi_i(x)$ are
functional derivatives.

This can also be derived using the Woodbury matrix identity (e.g. Press et al.,
1990). Using equations (\ref{lll}) and (\ref{llq}) we can write the
path-integral marginalized likelihood as
 \ba
\label{lll1}
    \calL \!\!\!\!&=& \!\!\!\!  \Delta \D\left( \C^{-1}
    - \C^{-1} \left[\int dx' dx''
      \frac{\delta \mub^\dag [\psi(x')]}{\delta  \psi_{i}(x')}
     M_{i j}^{-1}(x',x'')
     \frac{\delta \mub^\dag [\psi(x')]}{\delta  \psi_{j}(x')}
      \right]\C^{-1}\right)\Delta \D^\dag,
 \ea
where we have remove the constant term, and
\be
M_{ij}=F_{ij}+(C^{\psi}_{ij})^{-1}
\ee
where $F_{ij}$ is the path integral Fisher matrix given by
equation (\ref{llp}). Then using the Woodbury identity the covariance
term in equation (\ref{lll1}) can be directly written as equation
(\ref{lll0}).

\section*{Appendix B: Numerical Path Integration}

In this paper we make use of the assumption of Gaussianity in
function-space. However, in some cases this assumption will be a
poor approximation to the actual functional distribution. In this
case there are two alternative approaches which we review here:
Numerical Path-Integration and Functional Form-Filling.

\textbf{Numerical Path Integration} decomposes the
functional integration into discrete steps at position $x_i$, and
integrates over all possible field values at each position;
 \ba
L(\D|\thetab) &\propto&
\left[\prod_\alpha \int {\cal D} [\psi_{\alpha}(x)]\right]\,\,\,
p[\D|\thetab, \psib(x)]p[\psib(x)]
\nn
& = &
\left[\prod_\alpha  \,\, \prod_{i=1}^M \int\!d
\psi_\alpha(x_i)\right]\,\,\,
p[\D|\thetab, \psib(x_i)]p[\psib(x_i)] .
 \ea
Each integral can be evaluated by Monte-Carlo Markov-Chain (MCMC)
methods (e.g., Lewis \& Bridle 2002).

\textbf{Functional Form Filling} was developed in
Kitching et al., (2009) and uses a Monte-Carlo approach where the
bias in the measured parameters is evaluated instead of
marginalization. In the Appendices of Kitching et al (2009) it is
shown that the bias and marginalization procedures are equivalent.
Functional Form Filling is also a slight generalization of
Monte-Carlo path-integration in that arbitrary basis sets are used
and the coefficients of the expansion sampled randomly.

\section*{Appendix C : Path-Integral Marginalization for Power
  Spectrum Intrinsic Alignment Model}

Starting from Section \ref{Cosmic Shear Intrinsic Alignments}
equation (\ref{lensmean}) we can write the mean signal is the
cosmic shear and intrinsic alignment power spectra as
 \be
    \mu_{ij}(\ell)=C_{ij}(\ell)=C^{GG}_{ij}(\ell)+C^{GI}_{ij,j>i}(\ell)+N_{ij},
 \ee
where $i,j$ refer to the background source redshifts, $z_i$,
$z_j$, and we have intrinsic alignments for galaxies in the $i^{\rm th}$
redshift bin. The covariance of this is
\ba
\lefteqn{{\rm Cov} \left( [C^{GG}_{ij}(\ell)+C^{GI}_{ij,j>i}(\ell)]
  [C^{GG}_{lm}(\ell)+C^{GI}_{lm,m>l}(\ell)] \right)
  =} \nn &&
C^{GG}_{jm}(\ell) \left[ C^{GG}_{il}(\ell) + C^{GI}_{il,i>l}(\ell)
  + C^{GI}_{il,i<l}(\ell)\right] +
\left[C^{GG}_{im}(\ell)+C^{GI}_{im,i>m}(\ell) \right]
\left[C^{GG}_{jl}(\ell)+C^{GI}_{jl,l>j}(\ell)\right].
\ea
This includes all covariances between redshift bins pairs.

Starting from Section \ref{Cosmic Shear Intrinsic Alignments} we
can treat the ellipticity-density cross-power,
$P_{\gamma\delta}(\ell/r,r)$, itself as an unknown function.
Taking the functional derivative with respect to the perturbation
we find that
 \be
 \frac{\delta C^{GI}_{ij}(\ell)}{\delta P_{\gamma\delta}(\ell'/r,r)}
  =W^{GI}_{ij}(r) \frac{\delta_D(\ell-\ell')}{\ell}.
 \ee
In a similar way to Appendix B we can now write the marginalized
covariance as
 \ba
 \label{margC2}
    C^M_{\mu\nu}(\ell)=C_{\mu\nu}(\ell)+
    \int_0^{r_H} \!\! dr \,  \sigma^2_P(\ell/r,r) W^{GI}_{ij}(r)
    W^{GI}_{lm} (r),
 \ea
where $\sigma^2_P(\ell/r,r)=\lgl |\delta
P_{\gamma\delta}(\ell/r,r)|^2\rgl$ is the functional scatter in
$P_{\gamma\delta}$ and we have assumed the functional covariance
is diagonal in $\ell$ and $r$.

By taking the derivative of the mean with respect to a set
of cosmological parameters $\theta_{a}$ we can now write a
general expression for the Fisher matrix
\be
\label{margF}
F_{ab}=
\half \sum_{\mu \nu}\int \frac{\ell d \ell}{2 \pi}
\left([C^M_{\mu\nu}(\ell)]^{-1}
\left[\de_a \mu_{\mu}(\ell) \,
  \de_b \mu_{\nu}(\ell)
  +
  \de_a \mu_{\nu}(\ell) \,
  \de_b \mu_{\mu}(\ell) \right]\right)
\ee
where $\mu=(i,j)$ and $\nu=(l,m)$ denote pairs of background
redshifts. Equations (\ref{margC}) and (\ref{margF}) provide a
Fisher matrix for weak lensing tomography that include
marginalization over all possible GI intrinsic alignment bias
functions.

\section*{Appendix D: Path-Integral Marginalization for the Non-Linear Alignment Intrinsic
  Alignment Model}

In the following we will use the following  notation (similar to
Hirata \& Seljak, 2004 and Bridle \& King, 2007)
 \ba
    \label{com}
        C^{GG}_{ij}(\ell)=\int_0^{r_H} \! dr \, W^{GG}_{ij}(r)
        P_{\delta\delta}(\ell/r;r) \,\,\,\,
        {\rm and}
        \,\,\,\,
        C^{GI}_{ij}(\ell)=b^I(\ell)\int_0^{r_H} \! dr \, W^{GI}_{ij}(r)
        b^I(r)P_{\delta\delta}(\ell/r;r)
 \ea
which can be related to equations (\ref{GG}) and (\ref{GI}). We
have here introduced two bias functions $b^I(r)$ and
$b^I(\ell)$. We have assumed that the scale and redshift
dependent part of a more general bias $b^I(r,\ell)$ can be separated into
two independent functions in this case; this will allow us to
comment on the relative merit of constraining the bias function in
these directions.

In a similar way to Appendix C 
let us consider the bias functions as unknown systematics such
that we can write them as some fiducial function plus some
stochastic Gaussian distributed unknown $\psi_i(r)=\delta
b^I(r)$ and $\psi_j(\ell)=\delta b^I(\ell)$. We can write down
the functional derivatives associated with these perturbations as
 \ba
    \frac{\delta C^{GI}_{ij}(\ell)}{\delta b^I(r)}=
            b^I(\ell)W^{GI}_{ij}(r)P_{\delta\delta}(\ell/r;r)
    \,\,\,\,\,\,{\rm and}\,\,\,\,\,\,
    \frac{\delta C^{GI}_{ij}(\ell)}{\delta b_I(\ell')}=
        \frac{C_{ij}^{GI}(\ell)}{b_I(\ell)} \frac{\delta_D(\ell-\ell')}{\ell}.
 \ea
Using equation (\ref{com}) and assuming that the cross terms in
the nuisance function covariance matrix are zero, $ \lgl \delta
b_I(r) \delta b_I^*(\ell')\rgl=0$, and that the auto-covariance
are diagonal in $r$ and $\ell$ the marginalized covariance becomes
\ba \label{margC}
    C^M_{\mu\nu}(\ell)=C_{\mu\nu}(\ell)+
    b_I^2(\ell) \int_0^{r_H} \!\! d r \, \sigma^2_b(r)
    W^{GI}_{ij}(r) W^{GI}_{lm}(r) \left|P_{\delta\delta}(\ell/r;r)\right|^2
    +
    \left( \frac{\sigma_b^2(\ell)}{b_I^2(\ell)}\right)
    C^{GI}_{ij}(\ell) C^{GI}_{lm}(\ell),
 \ea
where $\lgl |\delta b_I(r)|^2\rgl=\sigma^2_b(r)$ and similarly for
$\ell$.

\end{document}